\documentclass{ws-procs9x6}

\begin{document}
\title{Neutron rich nuclei and neutron stars}

\author{C. J. Horowitz}%
       
\address{CEEM and Physics Department, Indiana University,\\ 
Bloomington, IN, 47405, USA \\
        E-mail: horowit@indiana.edu}


\begin{abstract}
The PREX experiment at Jefferson Laboratory measures the neutron radius of 208Pb with parity violating electron scattering in a way that is free from most strong interaction uncertainties.  The 208Pb radius has important implications for neutron rich matter and the structure of neutron stars.  We present first PREX results, describe future plans, and discuss a follow on measurement of the neutron radius of 48Ca.  We review radio and X-ray observations of neutron star masses and radii.  These constrain the equation of state (pressure versus density) of neutron rich matter.   We present a new energy functional that is simultaneously fit to both nuclear and neutron star properties.   In this approach, neutron star masses and radii constrain the energy of neutron matter.   This avoids having to rely on model dependent microscopic calculations of neutron matter.  The functional is then used to predict the location of the drip lines and the properties of very neutron rich heavy nuclei.
\end{abstract}

\bodymatter

\section{Introduction}
\begin{figure}[h]
\center\includegraphics[width=3.5in]{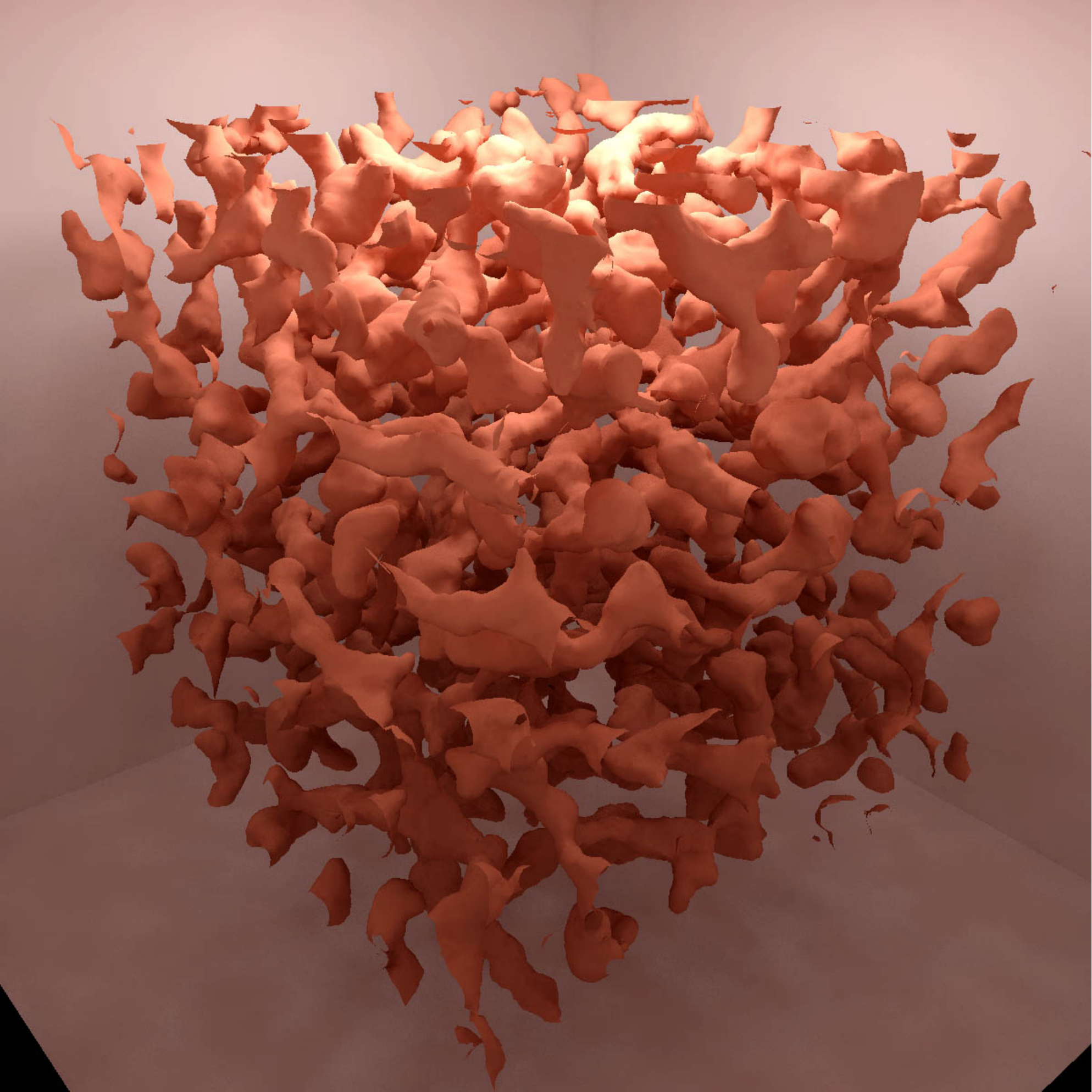}
\caption{\label{Fig1}Surfaces of proton density for a pasta configuration of neutron rich matter at a baryon density of 0.05 fm$^{-3}$.  This is from a semiclassical molecular dynamics simulation with 100,000 nucleons \cite{pasta2}. }
\end{figure}

Compress matter to high densities ($10^{11}+$ g/cm$^3$) and electrons are captured to make neutron rich matter. This material is at the heart of many fundamental questions in nuclear physics and astrophysics.
\begin{itemize}
\item What are the high density phases of QCD?
\item Where did the chemical elements come from?
\item What is the structure of many compact and energetic
objects in the heavens, and what determines their
electromagnetic, neutrino \cite{SNelastic}, and gravitational-wave \cite{GWbreaking}
radiations?
\end{itemize}
We are in Interested in neutron rich matter over a great range of density and temperature were it can be a gas, liquid, solid, plasma, liquid crystal (nuclear pasta), superconductor ($T_c=10^{10}$ K), superfluid, color superconductor.  The liquid crystal phases are known as nuclear pasta  \cite{pasta,watanabe}.  Pasta is expected at the base of the crust in a neutron star and can involve complex shapes such as long rods (``spaghetti'') or flat plates (``lasagna''), see Fig. \ref{Fig1}.   Pasta may scatter neutrinos in core collapse supernovae because it has sizes comparable to neutrino wave lengths \cite{pastascattering}.  

In Section \ref{sec.nuclei} we describe a precision laboratory experiment called PREX to measure the neutron radius of $^{208}$Pb.  In Section \ref{sec.tovmin} we describe a new energy density functional that is fit to properties of finite nuclei and the maximum mass and radius of neutron stars. We conclude in Section \ref{sec.conclusions}.

\section{The Lead Radius Experiment (PREX)}
\label{sec.nuclei}
Neutron rich matter can be studied in the laboratory.  We give one example, the Lead Radius Experiment (PREX) \cite{PREXI} accurately measures the neutron radius in $^{208}$Pb with parity violating electron scattering \cite{bigprex}.  This has important implications for nuclear structure, astrophysics, and atomic parity violation. 

\begin{figure}[ht]
\begin{center}
\includegraphics[width=3.7in,angle=0,clip=true] {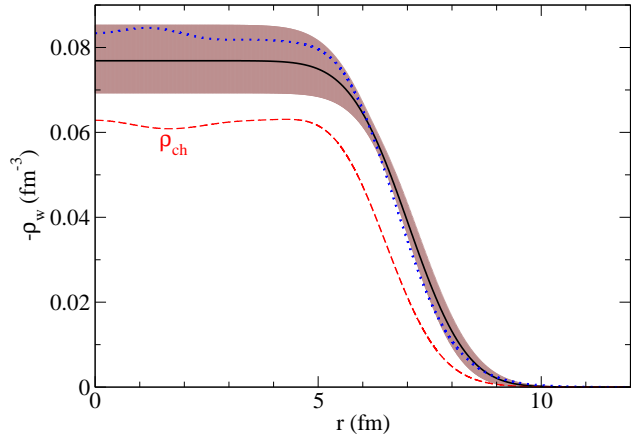}
\caption{Weak charge density $\rho_W(r)$ of $^{208}$Pb that is consistent with the PREX result (solid black line) \cite{weakFF}.  The brown error band shows the incoherent sum of experimental and model errors.   The red dashed curve is the experimental (electromagnetic) charge density $\rho_{ch}$.}
\label{fig:prex_den}
\end{center}
\end{figure}

Parity violating electron scattering provides a model independent probe of neutron densities that is free from most strong interaction uncertainties.  This is because the weak charge of a neutron is much larger than that of a proton \cite{dds}.  Therefore the $Z^0$ boson, that carries the weak force, couples primarily to neutrons.  In Born approximation, the parity violating asymmetry $A_{pv}$, the fractional difference in cross sections for positive and negative helicity electrons, is proportional to the weak form factor.  This is the Fourier transform of the weak charge density $\rho_W(r)$, see Fig. \ref{fig:prex_den}, and is closely related to the neutron density.  Therefore the neutron density can be extracted from an electro-weak measurement \cite{dds}.   Note that the Born approximation is not valid for a heavy nucleus and coulomb distortion effects must be included.  However, these have been accurately calculated \cite{couldist}.  
Many details of a practical parity violating experiment to measure neutron densities have been discussed in a long paper \cite{bigprex}.    

The Lead Radius Experiment (PREX) at Jefferson Laboratory \cite{PREXI} measures the parity violating asymmetry $A_{pv}$ for 1.05 GeV electrons elastically scattered from $^{208}$Pb at laboratory angles near five degrees.   The first measurement yielded $A_{pv}=0.656\pm 0.060$ (statistical) $\pm 0.014$ (systematic) ppm \cite{PREXI}.  From this the weak charge density has been deduced, see Fig. \ref{fig:prex_den}, and the rms neutron radius $R_n$ minus proton radius $R_p$ for $^{208}$Pb was found to be $R_n-R_p=0.33^{+0.16}_{-0.18}$ fm.  See ref. \cite{weakFF} for details of this analysis.  A second PREX run is now approved to accumulate more statistics and reach the original goal of determining $R_n$ to 1\% ($\pm 0.05$ fm).  

In addition to PREX, many other parity violating measurements of neutron densities are possible, see for example \cite{PREXII}.  Measuring $R_n$ in $^{48}$Ca is particularly attractive. With only 48 nucleons, microscopic coupled cluster calculations \cite{coupledcluster}, or no core shell model calculations \cite{NCSM}, are feasible for $^{48}$Ca that are presently not feasible for $^{208}$Pb.   Note that these microscopic calculations have important contributions from three nucleon forces.  This will allow one to make microscopic predictions for the neutron density and relate a measured $R_n$ to three nucleon forces. 

The neutron radius of $^{208}$Pb, $R_n$, has implications for astrophysics.  There is a strong correlation between $R_n$ and the pressure of neutron matter $P$ at densities near 0.1 fm$^{-3}$ (about 2/3 of nuclear density) \cite{alexbrown}.  A larger $P$ will push neutrons out against surface tension and increase $R_n$.  Therefore measuring $R_n$ constrains the equation of state (EOS) --- pressure as a function of density --- of neutron matter.  

Recently Hebeler et al. \cite{hebeler} used chiral perturbation theory to calculate the EOS of neutron matter including important contributions from very interesting three neutron forces.  We have some information on isospin 1/2 three nucleon forces from mass 3 nuclei ($^3$He, $^3$H) and proton-deuteron scattering.  However, our experimental information on three neutron forces is limited.  From their EOS, they predict $R_n-R_p= 0.17 \pm 0.03$ fm.  Here $R_p$ is the known proton radius of $^{208}$Pb.   Monte Carlo calculations by Carlson et al. also find sensitivity to three neutron forces \cite{MC3n}.   Therefore, measuring $R_n$ provides an important check of fundamental neutron matter calculations, and constrains three neutron forces.  
  
The EOS of neutron matter is closely related to the symmetry energy $S$.  
This describes how the energy of nuclear matter rises as one goes away from equal numbers of neutrons and protons.  There is a strong correlation between $R_n$ and the density dependence of the symmetry energy $dS/dn$, with $n$ the baryon density.  The symmetry energy can be probed in heavy ion collisions \cite{isospindif}.  For example, $dS/dn$ has been extracted from isospin diffusion data \cite{isospindif2} using a transport model.

The symmetry energy $S$ helps determine the composition of a neutron star.     A large $S$, at high density, implies a large proton fraction $Y_p$ that will allow the direct URCA process of rapid neutrino cooling.  If $R_n-R_p$ is large, it is likely that massive neutron stars will cool quickly by direct URCA  \cite{URCA}.  In addition, the transition density from solid neutron star crust to the liquid interior is strongly correlated with $R_n-R_p$ \cite{cjhjp_prl}.  

Finally, the correlation between $R_n$ and the radius of a neutron star $r_{NS}$, see for example Ref. \cite{SLB,Rutlidge}, is also very interesting \cite{rNSvsRn}.  In general, a larger $R_n$ implies a stiffer EOS, with a larger pressure, that will also suggest $r_{NS}$ is larger.  Note that this correlation is between objects that differ in size by 18 orders of magnitude from $R_n\approx 5.5$ fm to $r_{NS}\approx 10$ km.

\begin{figure}[ht]
\includegraphics[width=0.9\columnwidth,clip]{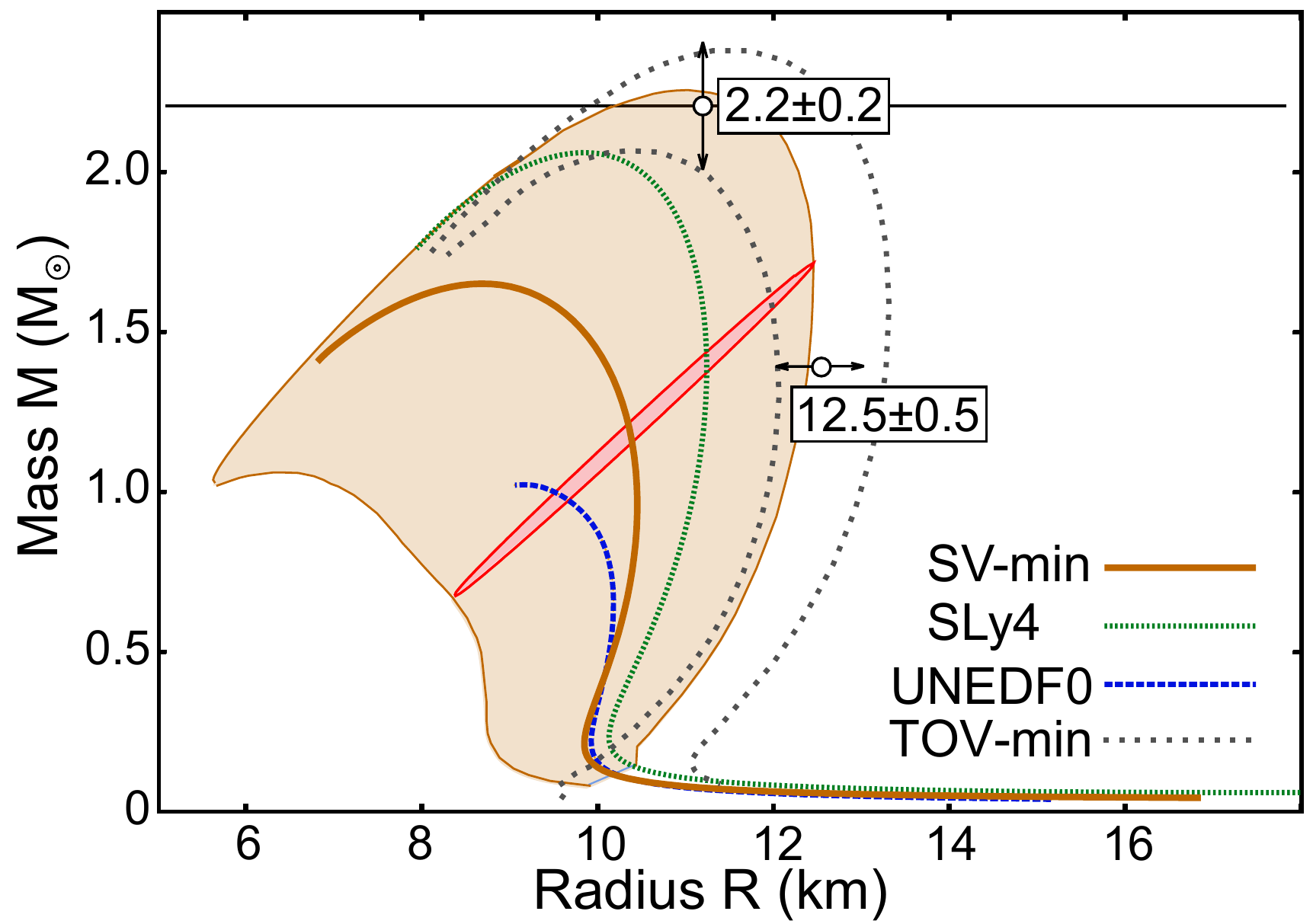}
\caption{Neutron star mass-radius relation of SLy4 \cite{SLy4}, UNEDF0 \cite{UNEDF0} and SV-min \cite{Klu09a}. The uncertainty band for SV-min is shown. This band is estimated by calculating the covariance ellipsoid for the mass $M$ and the radius $R$ at each point of the SV-min curve 
as indicated by the ellipsoid. Also depicted (dotted lines) are uncertainty limits for TOV-min.
}
\label{fig:uncert-SVmin}
\end{figure}

\section{Energy density functional for nuclei and neutron stars}
\label{sec.tovmin}

Energy density functionals (EDF) are often fit to properties of finite nuclei.  However, this can leave many properties of bulk neutron matter poorly constrained.  For example the SV-min functional has large uncertainties in its predictions of neutron star mass versus radius, see Fig. \ref{fig:uncert-SVmin}.   Alternative optimization protocols are  based on considering both experimental data on finite nuclei and theoretical pseudo-data on  nuclear and neutron matter. For example, the 
SLy4 \cite{SLy4} EDF was fit to both nuclei and the variational Fermi hypernetted chain  calculations of neutron matter \cite{FHNC}.  Chamel {\it et al.} calculated the maximum mass of neutron stars for three EOS fit to finite nuclei and microscopic neutron matter calculations \cite{Chamel:2011aa}, and such analysis helped them to so select one EDF (BSk21) that does well on both kinds of data.  

However, it is to be noted that current microscopic calculations of neutron matter with phenomenological interactions are highly model dependent; in particular, the impact of the poorly known three-neutron forces (especially their $T=3/2$ component) seems to be crucial \cite{Gan12a,Ste12a}. 
In this respect, calculations based on the chiral effective field theory 
with all many-neutron forces are predicted to N$^3$LO \cite{Heb10a,Hebeler10,Tews2013} can provide useful guidance, at least at sub-saturation densities. At high density, however, four or more-nucleon forces \cite{4Nforces,Tews2013} may also be important.  Indeed, at this time, our ability to calculate neutron matter properties at high densities is fundamentally limited.   

\begin{figure*}[htb]
\includegraphics[width=\textwidth]{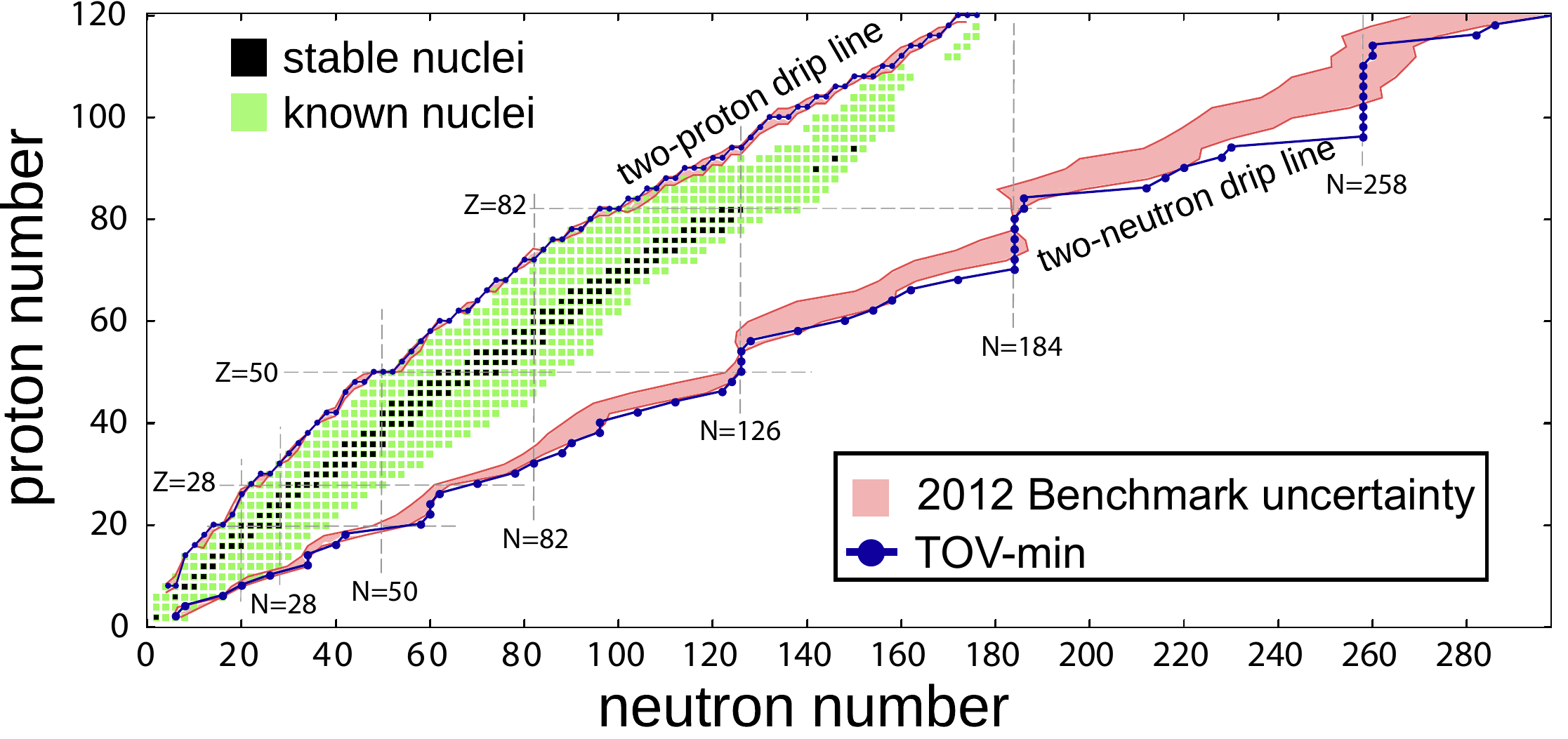}
\caption{\label{fig:drip}Two-nucleon drip line of TOV-min (dark blue) \cite{TOVmin} compared to 
the uncertainty band from Ref. \cite{Erl12a} (2012 Benchmark)
obtained by averaging results of six different 
EDF: SkM*, SkP, SLy4, SV-min, UNEDF0, and UNEDF1. The dashed grey grid lines show the magic numbers 
known around the valley of stability (20, 28, 50, 82, 126) as well as the predicted regions of enhanced
shell-stability in superheavy nuclei around $N=184$ and 258.  All 767 experimentally known even-even isotopes are shown with the stable nuclei indicated as black squares and the radioactive nuclei as green squares.
}
\end{figure*}

Recently we fit a new energy density functional, TOV-min, to both properties of finite nuclei and to the maximum mass of a neutron star $M_{max}$ and the radius of a 1.4 $M_\odot$ solar mass neutron star \cite{TOVmin} (see also ref. \cite{Fattoyev2010}).   This functional makes much more accurate predictions for neutron star mass versus radius, see Fig. \ref{fig:uncert-SVmin}.  The functional, with neutron matter properties constrained by neutron star data, then may make more accurate predictions for very neutron rich nuclei.  For example, Fig. \ref{fig:drip} shows the location of the drip line for a large range of nuclei.

\section{Conclusions: neutron rich nuclei and neutron stars}
\label{sec.conclusions}
Neutron rich matter is at the heart of many fundamental questions in Nuclear
Physics and Astrophysics. What are the high density phases of QCD? Where did the chemical elements come from? What is the structure of many compact and energetic objects in the heavens, and what determines their electromagnetic, neutrino, and gravitational-wave radiations? 

We described the Lead Radius Experiment (PREX) that uses parity violating electron scattering to measure the neutron radius in $^{208}$Pb. This has important implications for neutron stars and their crusts.  We also described a new energy functional fit to the properties of finite nuclei and to the masses and radii of neutron stars.  This functional then predicts properties for very neutron rich nuclei.

\section*{Acknowledgments}
Support for this work was provided in part through the Scientific Discovery through Advanced Computing (SciDAC) program funded by U.S. Department of Energy, Office of Science, Advanced Scientific Computing Research and Nuclear Physics, under award number DE-SC0008808 and by DOE grant DE-FG02-87ER40365.  Computer time was provided by the National Science Foundation, XSEDE grant TG-AST100014.  
\medskip
\bibliographystyle{plain}

\end{document}